%% file: SEtoka.tex
\documentclass{evn2004}
\usepackage{txfonts}
\usepackage{graphicx}
\begin{document}
\include{page}
   \title{MERLIN 6-GHz maser emission from W3(OH)}

   \author{S. Etoka\inst{1}
          \and
          R.J. Cohen\inst{1}
          \and
          M.D. Gray\inst{2}
          }

   \institute{Jodrell Bank Observatory, University of Manchester, 
              Macclesfield SK11 9DL
         \and
              Physics Department, UMIST, PO Box 88, Manchester M60 1QD}

   \abstract{
	We present the preliminary results of the first alignment to 
milliarsecond accuracy of the 6-GHz maser emission in OH and methanol 
maser lines toward W3(OH).
The identifications of Zeeman pairs allowed us to infer the actual velocity 
of the OH material and the magnetic field strength at the location where 
6.7-GHz methanol maser emission arises.
}
   \maketitle
%
%
\section{Introduction}
W3(OH) is a well-studied star forming region containing a Hot Molecular 
core (HMC) and an (UC)HII region (Turner \& Welch 1984). 
The central object is thought to be a O9-O7 young star 
with an estimated mass of about $30M_{\odot}$ (Dreher \& Welch 1981).
W3(OH) has been mapped in various maser lines: in the OH ground state lines
at 18~cm and in the OH excited lines at 4.7 and 6~GHz (Moran et al. 1978; 
Desmurs etal. 1998; Gray et al. 2001; Palmer et al. 2003; Wright et al. 
2004a,b), in the methanol lines at 6.7~GHz and 12~GHz 
(Menten et al. 1992; Moscadelli et al. 1999; Sutton et al. 2004).
The work presented here deals with MERLIN observations in the excited OH 
maser lines at 6031 and 6035~MHz and in the methanol maser line at 6668~MHz 
internally phased allowing accurate alignment of the three datasets and 
therefore a precise insight into copropagation. 
%
\section{Observations}
	Observations were performed on 12th March 2001 for all the 
lines. In the OH excited lines, observations were obtained at a rest 
frequency of 6030.747 and 6035.092-MHz for the J=5/2,F=2-2 and 3-3 
transitions respectively. Both main lines were recorded with a bandwidth 
of 0.5~MHz, divided into 256 channels. 
Observations in the 6668-MHz methanol line
were performed with a spectral bandwidth of 2~MHz, 
divided into 512 channels. 
The central velocity was taken to be $V_{LSR}=-45$~km.s$^{-1}$.
For all the three experiments 3C84 was used as the bandpass calibrator and 
0224+671 as a phase calibrator.
The initial data editing and the correction for gain-elevation effects 
were performed using the MERLIN dprogram packages. Further calibrations were 
performed within the AIPS package.   
%
\section{Results}
\begin{figure}
   \centering
   \includegraphics[width=7.5cm]{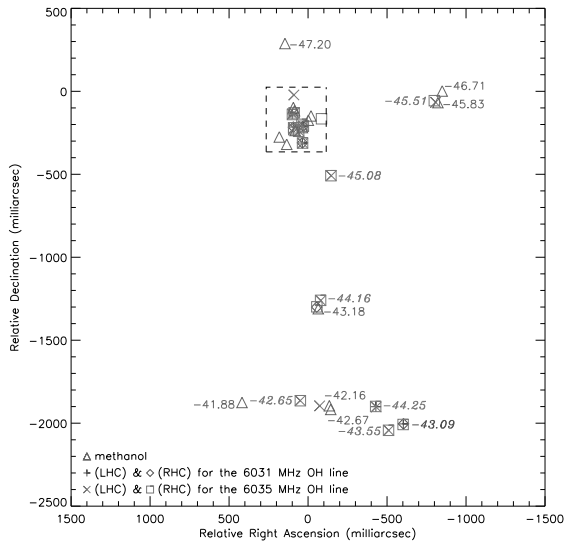}
   \caption{\small Maser feature distribution in the 6-GHz excited OH 
         lines and in the 6.7-GHz methanol line. Centre position (0,0) 
         corresponds to RA$_{\rm J2000}$=$02^h 27^m 03^s.820$, 
         Dec$_{\rm J2000}$=$61^o 52' 25.''400$. The velocities given for the 
	 excited OH lines in italic style are `demagnetized'
            \label{fig 1}
           }
    \end{figure}

\begin{figure}
   \centering
   \includegraphics[width=7.5cm]{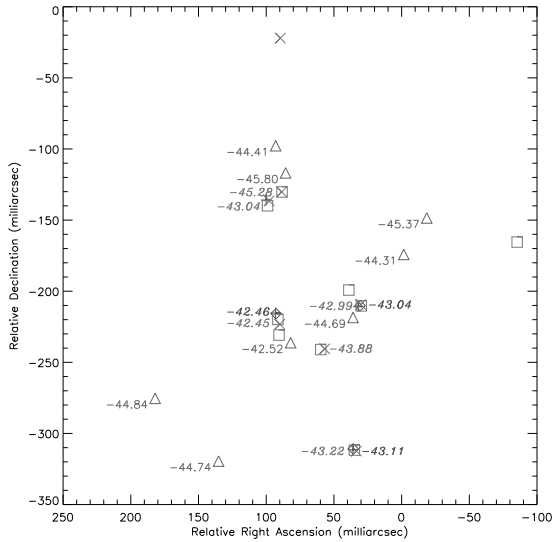}
      \caption{\small Magnification of the region given by the dotted box in 
	 Fig.~\ref{fig 1}
         \label{fig 2}
         }
   \end{figure}

Figures~\ref{fig 1} and~\ref{fig 2} present 
all the maser features detected in all the three sets of data. 
For the study of possible copropagation mechanisms, we determined the 
range for the FWHM from the spectra taken at the pixels where maser 
components were identified in the final images. The FWHM  was 
found to range typically from 0.26 to 0.30~km/s.
\subsection{Excited OH at 6035~MHz}
We detected 31 components (14 LHC and 17 RHC) at that frequency.
Twelve possible Zeeman pairs were identified. Eleven of them are 
identified with a separation of less than 5 mas while one pair shows a 
separation of about 10~mas between the two pairs. All the pairs but one are 
such that $V_{\rm LHC} < V_{\rm RHC}$.
One Zeeman pair is such that $V_{\rm LHC} \ge V_{\rm RHC}$. 
This may be due to blending of multiple components since it is known 
that this area contains a large number of maser spots with a wide range 
of velocity. 
\subsection{Excited OH at 6031~MHz}
We found 11 components (6 LHC and 5 RHC) 
from which 4 Zeeman pairs could be identified. 
Three out the four Zeeman pairs observed at 6031-MHz were found coincident 
with 6035-MHz Zeeman pairs. The very good agreement in position ($<6.5$~mas) 
and demagnetized velocity ($\Delta V \leq 0.11 <$FWHM) are such that 
we are fairly confident that we are observing copropagating masers. 
It is also noteworthy that all 6031-MHz spots 
coincide with 6035-MHz maser spots within 6.5~mas.
\subsection{Methanol at 6668~MHz}
We found 15 components.
Five methanol maser spots (i.e., 33\%) are found within 20~mas from an OH 
6035-MHz maser spot. Three of these maser spots belong to the crowded area 
RA$_{\rm offset}$=[+250,-100]~mas, Dec$_{\rm offset}$=[0,$-350$]~mas, 
one belongs to the far West area RA$_{\rm offset}$=[-800,-850]~mas, 
Dec$_{\rm offset}$=[0,-100]~mas and the last one to the area 
RA$_{\rm offset}$=[+100,-200]~mas, Dec$_{\rm offset}$=[-1200,-1400]~mas.
The maser spot belonging to the last mentioned southern area and one 
belonging to the crowded area RA$_{\rm offset}$=[+250,-100]~mas, 
Dec$_{\rm offset}$=[0,-350]~mas
are such that the velocity separation between the methanol maser spot and 
the demagnetized velocity of the nearby OH Zeeman pair is about 1~km/s. 
This is far bigger than 2*FWHM. It is therefore quite unlikely that those 
maser signatures actually arise from the same body of gas. 
On the other hand, for the three other `quasi coincident' maser spots in 
both species the velocity difference ranges from 0.07~km/s to 0.52~km/s, 
that is $\le$2*FWHM. It is therefore quite probable that they belong to the 
same body of gas.
%

%
\section{Findings}
\subsection{Magnetic field}
From the Zeeman pairs identified for the excited OH 6-GHz datasets we 
calculated the magnetic field strength over the whole W3(OH) star forming 
region . This latter has been found to range from a value of 0.4~mG to 14.5~mG 
in agreement with the previous study of Desmurs et al. (1998). 
The tight assocations ($<15$~mas) between methanol maser spots and OH 
maser Zeeman spots at 6-GHz, for which the velocity agree well for the 2 
species to be possibly part of the same body of gas, allowed us a probe of 
the magnetic field strength at the location of methanol masing regions.
Interestingly, this investigation revealed a rather strong magnetic 
field ($\ge$7~mG) at such location. 
\subsection{Velocity field and density}
A velocity gradient is clearly observed in what is thought to be a 
circumstellar disk seen edge-on delimited roughly by 
RA$_{\rm offset}$=[-150;-150]~mas and Dec$_{\rm offset}$=[+300;-1500]~mas. 
The southern area at Dec$_{\rm offset}$=[-1800,-2100]~mas 
is such that the East area is red shifted while the West area is blue shifted 
with respect to the velocity of the southern extremity of the disk. 
This area is thought to be tracing a shock wave.
For a better picture of the physical conditions at the various locations of 
maser emission found in W3(OH), we aligned our 6-7-GHz dataset to that of 
Wright et al. (2004a,b) presenting the OH ground state maser distribution 
in W3(OH). Doing so, we found that interestingly, the southern area produces 
methanol maser emission, both excited OH maser emission at 6031 and 6035-MHz 
and all the ground state OH maser lines but the 1720-MHz line. According to 
the models developed by Gray et al. (1992), a strong production of 6031 and 
6035-MHz with a substantially fainter 1720-MHz is only acheived for relatively 
high densities ([H$_2$]=2.5 10$^7$~cm$^{-3}$ and [OH]=250~cm$^{-3}$) and the
lower end of the kinematic temperatures (T$_k$=75~K) allowing strong OH 6-GHz 
inversion. 
Also, Gray et al. mention the importance of collision in the 6035-MHz 
inversion. 
The lack of 1720-MHz with a strong production of 6-GHz OH excited 
maser strenghen the hypothesis of a shock wave as an explanation for the 
southern maser band.
%
\section{Conclusion}
	We have presented the first alignment to milliarsecond accuracy of 
the 6-GHz maser emission in OH and methanol maser lines toward W3(OH). It has
been found that all 6031-MHz spots are associated with a 6035-MHz spot within 
less than 6.5~mas separation while only 33\% of the methanol maser 
spots are found within 20~mas from a 6-GHz OH maser spot, none of them within 
less than 10~mas.
%

\end{document}

%% file: page.tex
\setcounter{page}{199}